\documentclass[%
 aip,
 amsmath,amssymb,
 reprint,%
]{revtex4-1}

\usepackage{graphicx}
\usepackage{dcolumn}
\usepackage{bm}

\usepackage[utf8]{inputenc}
\usepackage[T1]{fontenc}
\usepackage{mathptmx}
\usepackage{etoolbox}

\makeatletter
\def\@email#1#2{%
 \endgroup
 \patchcmd{\titleblock@produce}
  {\frontmatter@RRAPformat}
  {\frontmatter@RRAPformat{\produce@RRAP{*#1\href{mailto:#2}{#2}}}\frontmatter@RRAPformat}
  {}{}
}%
\makeatother
\begin{document}

\preprint{AIP/123-QED}

\title{Waveguide-Plasmon Polariton Quasiparticles with Exceptional Point Characteristics}
\author{P. Chang}
\email{pohan.chang@mail.utoronto.ca} 
 \affiliation{The Edward S. Rogers Sr. Department of Electrical and Computer Engineering, University of Toronto, 10 King’s College Road, Toronto, Ontario M5S 3G4, Canada}

\author{S. Ramezanpour}%
 \email{shahab.ramezanpour@utoronto.ca}
\affiliation{The Edward S. Rogers Sr. Department of Electrical and Computer Engineering, University of Toronto, 10 King’s College Road, Toronto, Ontario M5S 3G4, Canada}

\author{A. Helmy}
 \email{a.helmy@utoronto.ca}
\affiliation{The Edward S. Rogers Sr. Department of Electrical and Computer Engineering, University of Toronto, 10 King’s College Road, Toronto, Ontario M5S 3G4, Canada}

\date{\today}

\begin{abstract}
The growing complexity of integrated photonics necessitates compact, low-power devices that transcend traditional, material-centric design approaches. In this study, we harness non-Hermitian physics to uncover novel properties of coupled plasmonic waveguide modes exhibiting exceptional point (EP) degeneracy. Our hybrid plasmonic waveguide architecture, capable of supporting both strong and weak coupling regimes between plasmonic and dielectric waveguide modes, is precisely engineered to reach an EP where eigenmodes coalesce. This strategic tuning not only enhances the modal contrast between minimized-loss and highly dissipative states but also enables unprecedented control over device characteristics. Our findings introduce a new paradigm in integrated photonics, paving the way for ultracompact modulators and highly tunable on-chip communication systems with reduced power consumption.
\end{abstract}

\maketitle

\section{Introduction}
	The interaction between light and matter is fundamental in both science and technology, underpinning the operation of diverse optoelectronic devices. Light can couple to highly confined electromagnetic fields, including plasmonic, phononic, excitonic, and magnonic fields. The resemblance of these fields to photons motivates their classification as photonic quasiparticles—quantized solutions to Maxwell’s equations in a medium. However, these quasiparticles differ significantly from photons in free space, particularly in polarization, confinement, and dispersion \cite{Q1,q2}. Strong coupling between localized particle plasmons and optical waveguide modes drastically alters transmission spectra, leading to the emergence of a new quasiparticle—a waveguide-plasmon polariton—characterized by substantial Rabi splitting \cite{q3,q4}. Despite these intriguing properties, the behavior of coupled surface plasmon polariton (SPP) and waveguide modes as a non-Hermitian system remains largely unexplored.
	
	Since the discovery of real eigenfrequencies in non-Hermitian Hamiltonians that commute with parity-time (PT) operators \cite{A1}, PT symmetry and its associated spectral degeneracy—termed an Exceptional Point (EP)—have attracted increasing interest beyond mathematical physics. PT-symmetric systems remain invariant under simultaneous parity (P) and time-reversal (T) transformations. While inherently non-Hermitian, these systems can exhibit entirely real eigenvalues. Beyond a critical threshold, however, their eigenvalues transition into the complex domain (broken PT symmetry), exhibiting the defining characteristics of an EP. An EP is a degeneracy in an open system where both eigenvalues and their corresponding eigenmodes coalesce. Initially introduced in quantum mechanics, PT symmetry has been successfully realized in optics through symmetric index guiding combined with an antisymmetric gain/loss profile, as observed in coupled two-component systems \cite{A2}. Advances in PT-symmetric photonic structures \cite{A3,8,9,10} have facilitated numerous nontrivial effects, including PT-symmetric lasers \cite{A4,11}, enhanced sensing \cite{12,13,14}, chiral and directional propagation \cite{15,16}, robust wireless power transfer \cite{17}, unidirectional invisibility \cite{18}, coherent perfect absorption \cite{19,20}, and plasmonic nanoscale sensing \cite{21}. Furthermore, incorporating nonlinearity in PT-symmetric systems \cite{22,ramezanpour2024dynamic,23} significantly expands their tunability.
	
	PT symmetry and EPs can also be realized within a single component and between different cavity modes \cite{A5,A6,ramezanpour2021generalization,ramezanpour2024highly}, as well as in guided modes with distinct polarizations \cite{doppler2016dynamically,A7}, through structural deformations. In this context, dynamical encircling of an exceptional point (EP) — i.e., varying system parameters along a closed path that encloses an EP in parameter space — produces striking nonreciprocal and chiral state-transfer phenomena: slow (quasi-adiabatic) encirclement can induce a state flip and a nontrivial geometric phase, while the breakdown of adiabaticity for other encirclement protocols leads to robust, direction-dependent mode conversion and asymmetric transmission. Recent theoretical and experimental work has exploited these effects for polarization-controlled and chiral transport, fast encirclement schemes that realize highly efficient, compact mode converters, and Riemann-encircling strategies that improve conversion fidelity and device compactness \cite{zhu2025polarization,shu2024chiral,shu2022fast,li2022riemann}.
    
    These advancements, coupled with previous progress in PT-symmetric systems, have motivated further investigation into non-Hermitian degeneracies among modes with fundamentally different characteristics. For instance, EPs have been observed in coupled systems involving graphene plasmons and vibrational modes \cite{A8}, in cavity magnon-polaritons \cite{A9,lambert2025coherent}. Additionally, hybridization between magnons and plasmons has been demonstrated in a heterostructure of a graphene layer and a 2D ferromagnetic layer \cite{A10}.
	
	A hybrid plasmonic waveguide mode merges the properties of dielectric and plasmonic waveguides to achieve superior performance. In these structures, light is confined at the subwavelength scale by the plasmonic component (typically a metal), while the dielectric component ensures lower propagation loss over extended distances. This synergy enables strong optical confinement, enhanced field interactions, and seamless integration with conventional photonic circuits, making hybrid plasmonic waveguides well-suited for nanoscale optics, sensing, and integrated photonics \cite{1,2,Mo,3,4,5,6,7}. A composite hybrid plasmonic waveguide (CHPW) structure, incorporating a low-index dielectric such as $SiO_2$ beneath a thin metallic film and sandwiched between semiconductor waveguides, has demonstrated applications in Si-based integrated photonics, including highly sensitive plasmonic detectors \cite{5} and record Purcell factor plasmonic ring resonators \cite{6}.
	
	Recently, our group has verified significant modal overlap between hybrid plasmonic and waveguide modes through precise parameter space engineering \cite{7}. The interaction between the lossy short-range SPP mode and long-range waveguide mode can be effectively described using a non-Hermitian Hamiltonian. A key characteristic of non-Hermitian systems is the presence of exceptional points (EPs), defined by the coalescence of both eigenvalues and eigenvectors. This study demonstrates that, through careful structural design, EPs can be realized in hybrid plasmonic waveguides. This advancement holds both fundamental significance—by enabling the observation of a quasiparticle with unique properties—and practical implications for ultra-compact modulation, detection, and on-chip optical circuits. By leveraging 2D materials with anisotropic \cite{24,25}, nonlinear \cite{26}, and tunable \cite{27,28,29} properties, it becomes possible to engineer devices with electrically tunable EPs at the nanoscale. 
    
    We believe this extraordinary behavior could significantly impact plasmonic nanophotonics, leading to highly tunable, ultracompact on-chip components compatible with integrated photonics: Designs that balance confinement, loss, and manufacturability—delivering compact footprints, low drive voltages, and CMOS compatibility—are especially valuable for practical, large-scale photonic integration \cite{eppenberger2023resonant,salamin2018100,ummethala2021hybrid}.
\section{Design}
Any quantum system connected to an environment is considered an open system, described by a non-Hermitian effective Hamiltonian $\hat{H}_{0} \ne \hat{H}_{0}^{\dagger}$. If the Hamiltonian is non-normal, i.e., the Hamiltonian does not commute with its Hermitian conjugate, $\left[ \hat{H}_{0}, \hat{H}_{0}^{\dagger} \right] \ne 0$, the (right) eigenstates of a Hamiltonian are often mutually non-orthogonal. Near exceptional points (EPs) in parameter space, where at least two eigenstates become collinear, the nonorthogonality can be very significant \cite{0}. Realization of Exceptional Points in plasmonics at the nanoscale is intriguing due to their applications in silicon photonics, such as ultracompact modulators. Although being non-Hermitian (open systems), creating non-orthogonal modes in plasmonic waveguides is challenging.

Specifically, we demonstrate that the CHPW structure can be considered a photonic molecule with confined modes containing uncoupled short-range/long-range modes. At specific spatial parameters, the loss of the short-range (long-range mode) is substantially reduced (increased), leading to degenerate modes with the characteristics of an Exceptional Point (EP) (Fig. 1).

\begin{figure*}[ht!]
 	\centering\includegraphics[width=15cm]{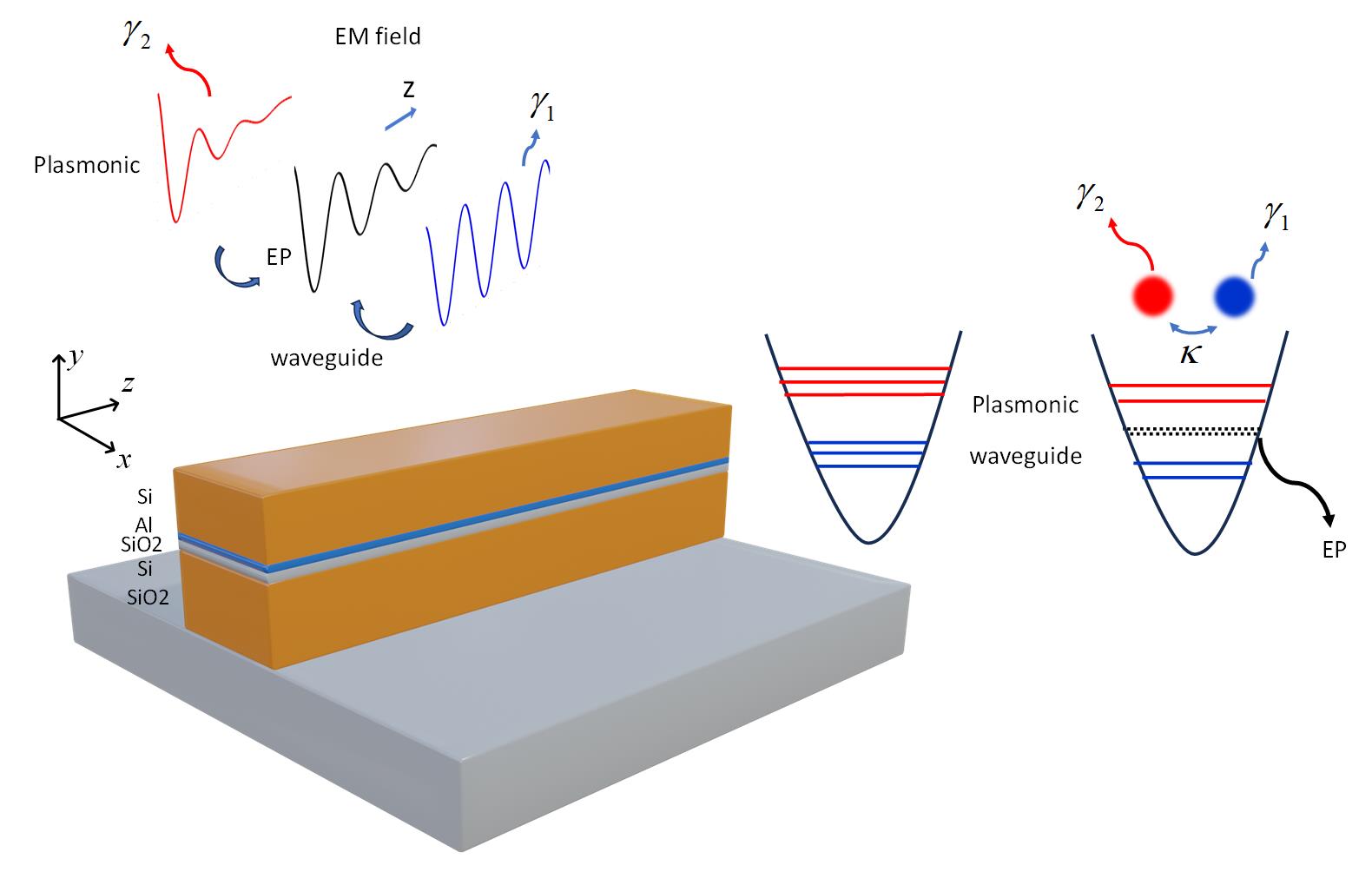}
	\caption{Observation of exceptional point in hybrid plasmonic waveguide. The short-range and long-range modes have the characteristics of plasmonic and waveguide modes, respectively. The confined modes can be considered as confined energy levels in a potential well as in photonic molecules. The modes inherently do not overlap with each other. However, with judicious tuning of space parameters, the short-range and long-range modes can overlap, and in extreme cases, these two modes coalesce (EP).}   
\end{figure*}

In the CHPW structure, the coupling of SPPs from both sides of the metallic thin layer and the coupling of SPPs to the waveguide mode through the low-index dielectric thin layer create additional degrees of freedom to engineer the modes' dispersion. With non-Hermitian properties, spectral degeneracy is achieved in this hybrid plasmonic structure through fine-tuning of the parameter space. To achieve the EP condition, both the real and imaginary parts of the effective refractive indices of the two modes should degenerate, and the difference in their imaginary parts should be twice the coupling rate between the modes. In particular, the $SiO_2$ thin layer, ridge width, and thickness of the top Si layer mainly affect the coupling between the modes and the real and imaginary parts of the effective indices, respectively. As shown schematically in Fig. 2(a), the width of the layer structure is optimized to create an overlap between two distinct plasmonic and waveguide modes with different losses and field distributions (the field distribution shown by the red color corresponds to $|E_y|$). 

The schematic adiabatic taper is one practical method to selectively excite the fundamental slot mode: by gradual geometry conversion it enforces phase-matching, preserves TM polarization, and maximizes spatial overlap with the slot field while suppressing scattering into higher-order and TE modes. For the present study we therefore focus on 2D cross-section eigenmode simulations and systematic width sweeps to map the complex modal indices and exceptional-point behaviour.

In a metal thin layer sandwiched between two identical dielectric slabs, the bound Surface Plasmon Polaritons (SPPs) at the top and bottom of the metal-dielectric interfaces couple to form two TM-polarized bound supermodes, denoted asymmetric and symmetric bound \cite{r1}. Let us initially consider a plasmonic waveguide structure, as schematically shown in Fig. 2(b), which consists of a metal film between two Si layers with finite widths. Using the Finite Difference Eigenmode (FDE) solver in Lumerical, our numerical results show that this structure also has modes with (nearly) asymmetric and symmetric field (real part of $E_y$) distribution analogous to its counterpart with infinite widths. All simulations were performed at a wavelength of $1.55\ \mu\mathrm{m}$, chosen because it lies in the standard telecom band where silicon exhibits low loss, material optical constants are well characterized, and mature fabrication/measurement infrastructure (CMOS-compatible processes, laser sources, fiber coupling and detectors) makes the results directly relevant to practical integrated-photonics implementations.

Figures 2(c) and 2(d) show (real) $n_{eff}$ of the supermodes and their associated propagation loss, respectively, vs. the width of the layers. The modes with lower and higher imaginary parts of the (effective) refractive index are denoted as $TM_{LR}$ and $TM_{SR}$, respectively \cite{30,5}. As demonstrated in Fig. 2(d), the propagation loss of the fundamental $TM_{SR}$ supermode is two orders of magnitude higher than that of $TM_{LR}$. The higher (lower) loss in $TM_{SR}$ ($TM_{LR}$) can be attributed to the interference of SPPs at the top and bottom of the metal interfaces, which is constructive (destructive), leading to a higher (lower) overlap of the field, especially the longitudinal component, $E_z$, with the metal thin layer.

\begin{figure*}[ht!]
	\centering\includegraphics[width=16cm]{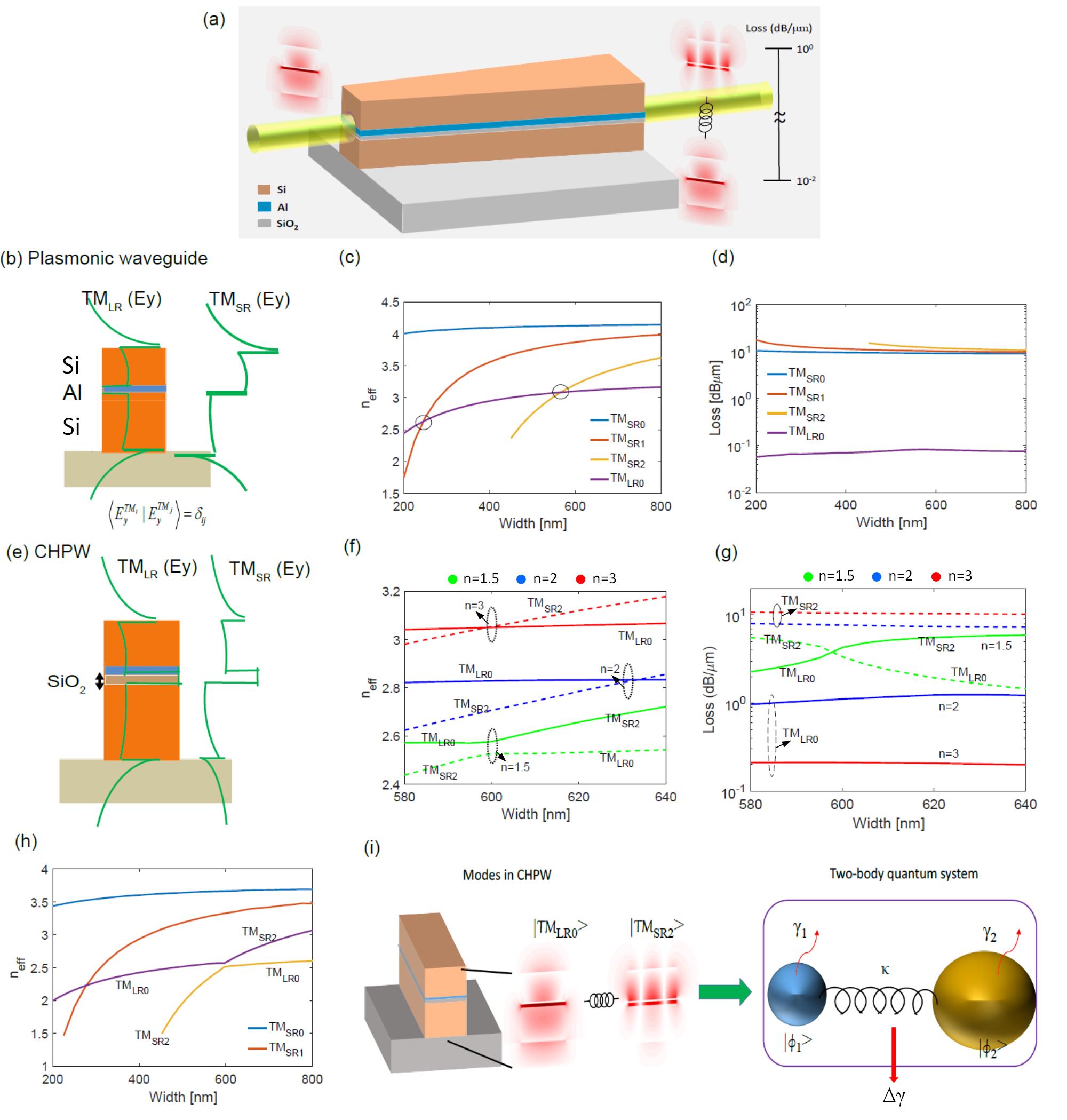}
	\caption{(a) A layered structure enables the overlapping between waveguide and plasmonic modes through changes in the width of the layers. (b) A plasmonic waveguide with an $Al$ film between two Si layers. (c) Effective refractive indices and (d) propagation loss of the supermodes. (e) Structure with an $SiO_2$ thin layer beneath an $Al$ thin layer. (f) $n_{eff}$ and (g) propagation loss of the modes in the structure for different values of the refractive index of the thin layer beneath the metal layer. (h) Achieving degenerate modes in the structure in parameter space. (i) Quantum mechanical approach to model the evolution of $TM_{LR0}$ ($\phi_1$) and $TM_{SR2}$ ($\phi_2$) modes in the structure.}
\end{figure*}

By changing the width of the layers, level crossing occurs in the (real part) of the effective refractive indices. As demonstrated in Fig. 2(c), this crossing is observed between the $TM_{LR0}$ and $TM_{SR1}$ modes, and between the $TM_{LR0}$ and $TM_{SR2}$ modes. In other words, we may mention that changing the width enables phase-matching between the supermodes. However, at the crossing points, the modes are still orthogonal, and their overlapping is zero, $\left\langle E_{y}^{TM_{i}} | E_{y}^{TM_{j}} \right\rangle = \delta_{ij}$.

To create an overlap between the modes, a thin layer with refractive index $n$ is placed beneath the metal thin layer (Fig. 2(e)). For different values of $n$, Figs. 2(f) and 2(g) show the (real part of) $n_{eff}$ and propagation loss (related to the imaginary part of $n_{eff}$) vs. the width. These figures clearly show that for $n=3$ and $n=2$, level crossing occurs in $n_{eff}$ for specific values of widths, while at the same widths, level repulsion happens in the loss. Conversely, for $n=1.5$, level repulsion and crossing occur in $n_{eff}$ and loss, respectively. This level crossing and repulsion indicate the potential of the structure to encircle the exceptional point in parameter space.

Figure 2(h) depicts $n_{eff}$ for the structure with an $SiO_2$ thin layer beneath an $Al$ layer, where with precise design of the parameters, a degenerate point between $TM_{LR0}$ and $TM_{SR2}$ is obtained at a width of around $598$ nm. The $SiO_2$ thin layer enables the coupling between the $TM_{LR0}$ and $TM_{SR2}$ modes, which we may refer to as hybrid waveguide and plasmonic modes, respectively \cite{5,6,31}. In this hybrid structure, we can change the coupling and overlap between the modes by adjusting the thickness and width of the layers. If we model the structure as a two-level system containing $TM_{LR0}$ and $TM_{SR2}$ modes, by tuning the parameters, the system experiences strong and weak coupling regimes, and in an extreme case, when a balance between coupling and loss difference between the modes occurs, one can reach an exceptional point (Fig. 2(i)). See Supplementary Information Sec.~I for additional details on the modal dynamics near the exceptional point and on the occurrence of multiple guided modes.

\section{Results}
In a CHPW structure, one can create coupled modes with tunable dissipation and coupling rates by changing the dimensions of the layers. This platform, shown schematically in Fig. 3(a), enables achieving EP without inducing material loss or gain, at the nanoscale range. For the hybrid waveguide plasmonic modes, $TM_{LR0}$ and $TM_{SR2}$, one can model the structure by a two-level system with the Hamiltonian:

\begin{equation}
	H=\left( \begin{matrix}
		{{n}_{1}}-i{{\gamma }_{1}} & \kappa   \\
		{{\kappa }^{*}} & {{n}_{2}}-i{{\gamma }_{2}}  \\
	\end{matrix} \right),
\end{equation}

where ${{n}_{1}}-i{{\gamma }_{1}}$ and ${{n}_{2}}-i{{\gamma }_{2}}$ are related to the refractive indices of the waveguide and plasmonic modes, respectively, $\kappa$ is the coupling rate between these modes, and $\kappa^*$ is the complex conjugate of $\kappa$. The eigenvalues of the Hamiltonian can then be found as:

\begin{equation}
	n=\frac{{{n}_{1}}+{{n}_{2}}-i({{\gamma }_{1}}+{{\gamma }_{2}})\pm \sqrt{{{\left( {{n}_{1}}-{{n}_{2}}-i({{\gamma }_{1}}-{{\gamma }_{2}}) \right)}^{2}}+4|\kappa {{|}^{2}}}}{2}.
\end{equation}

To have EP, the term $\sqrt{{{\left( {{n}_{1}}-{{n}_{2}}-i({{\gamma }_{1}}-{{\gamma }_{2}}) \right)}^{2}}+4|\kappa {{|}^{2}}}=0$, which gives ${{n}_{1}}={{n}_{2}}={{n}_{0}}$ and $\Delta \gamma =\kappa$, where $\Delta \gamma=({{\gamma }_{1}}-{{\gamma }_{2}})/2$, and $\kappa$ is assumed a real value. See Supplementary Information Sec.~II for details of the coupled-mode theory and the sensitivity analysis near the exceptional point.

For further analysis, Figs. 3(b)-(c) show the refractive index and loss of the coupled modes vs. the width of the layers and for the thickness of the top silicon layer $h=250$, $h=305$ nm, and $h=330$ nm, respectively, which clearly reveal the transition of the eigenvalues from weak coupling to strong coupling regimes in parameter space. For $h=305$ nm and a width of around $598$ nm, both the refractive index and propagation loss degenerate, which is the EP condition.

\begin{figure*}
	\centering\includegraphics[width=17cm]{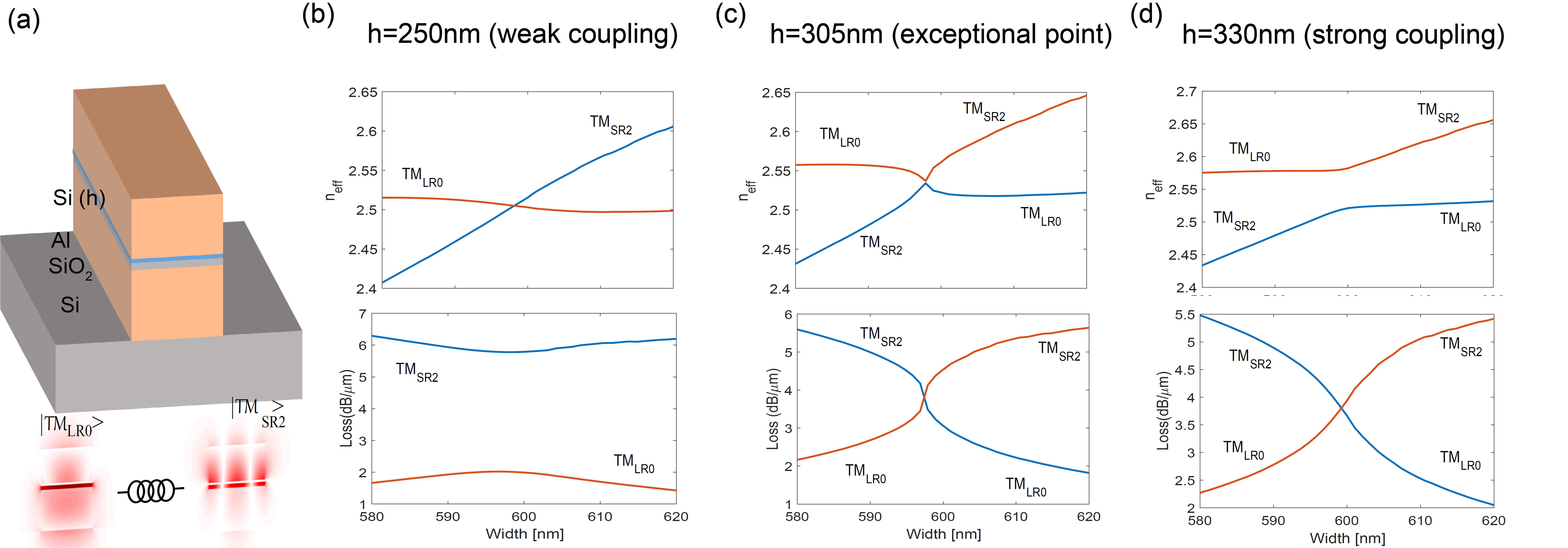}
	\caption{(a) By tuning the thickness of the top Si layer, the mode evolution in the structure reveals (b) weak coupling regime, (c) exceptional point, and (d) strong coupling regime.}
\end{figure*}

Finally, Fig. 4 shows the overlap between the modes with the total electric field $\psi_{TM_{LR0}}$ and $\psi_{TM_{SR2}}$, defined by parameter $F=\left\langle {{\psi }_{LR0}} | {{\psi }_{SR2}} \right\rangle$ vs. the thickness of the top silicon layer, calculated for the layer width around $598$ nm. At EP, the overlap is maximum, and its imaginary part changes sign on either side of the EP, which fulfills the definition of the EP that both eigenvalues and their corresponding eigenmodes should coalesce.
The degenerate modal profiles are shown in Supplementary Information Sec.~I.
\begin{figure}
	\centering\includegraphics[width=9cm]{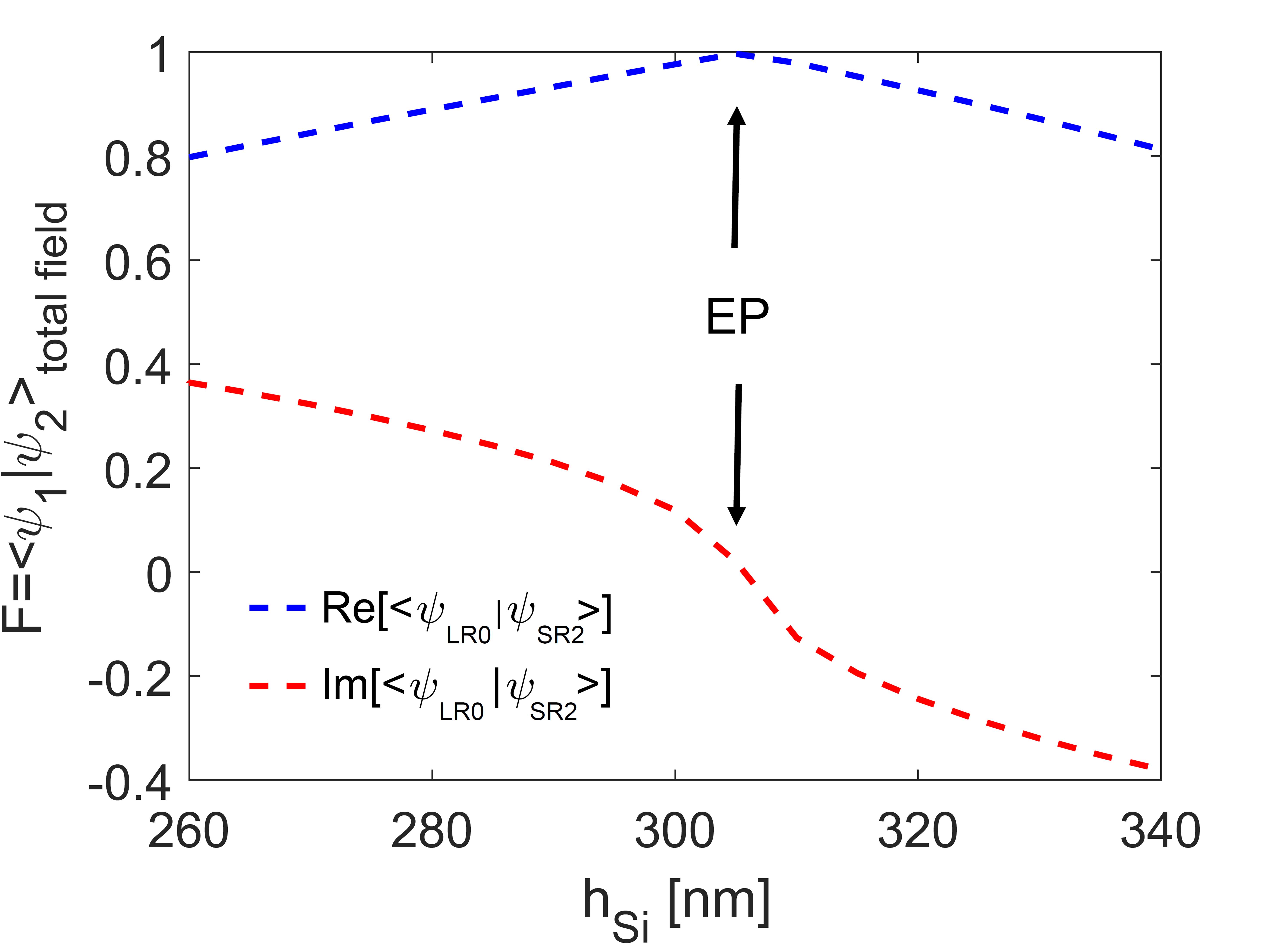}
	\caption{Overlap between the modes vs the height of the top Si layer, for the width of layer $598$ nm.}
\end{figure}

	\section{Discussion}
	Our study delves into the interplay of non-Hermitian physics and hybrid plasmonic waveguides, revealing unprecedented behavior of waveguide-plasmon polariton quasiparticles at exceptional points (EPs). Building on the seminal discovery that non-Hermitian Hamiltonians with PT-symmetry can exhibit real eigenfrequencies, our work demonstrates that careful engineering of coupling between plasmonic and dielectric waveguide modes can be harnessed to drive the system to an EP. At this critical juncture, the eigenvalues and eigenmodes coalesce, marking a transition between regimes characterized by starkly contrasting loss profiles. This behavior is emblematic of the sensitive balance between the short-range plasmonic mode and long-range dielectric waveguide mode, whose interaction gives rise to a hybrid quasiparticle with tunable propagation characteristics.
	
	The achievement of EP degeneracy in our composite hybrid plasmonic waveguide structure is nontrivial, requiring precise control over geometrical parameters such as layer thickness and width, as well as the intrinsic material properties. By manipulating these parameters, we navigate the system’s parameter space to identify regimes of both strong and weak coupling. In the strong coupling regime, the formation of waveguide-plasmon polaritons is observed, characterized by deep subwavelength confinement and significant field enhancement. Importantly, the transition through the EP facilitates a marked divergence between long-range and short-range modes, enabling ultracompact modulation capabilities with minimal power consumption.
	
	Our rigorous theoretical models, supported by state-of-the-art numerical simulations, underscore the pivotal role of non-Hermitian degeneracies in modulating the system's optical response. The sensitivity of the EP to parameter variations not only enhances our understanding of modal dynamics in hybrid systems but also paves the way for novel applications in integrated photonics. The tunability of the EP, achieved through the incorporation of advanced materials such as epsilon-near-zero (ENZ) layers and nonlinear media, offers promising avenues for the development of dynamically reconfigurable photonic devices.
	
	\section{Conclusion}
	In conclusion, our investigation establishes that hybrid plasmonic waveguides can be engineered to exhibit exceptional point degeneracies, thereby merging the benefits of strong light–matter interaction with the operational advantages of long-range dielectric waveguides. The emergence of waveguide-plasmon polariton quasiparticles at the EP not only enriches the fundamental understanding of non-Hermitian photonics but also heralds a new paradigm for the design of ultracompact, low-power photonic devices. The demonstrated ability to precisely tune the system's modal characteristics—via both structural design and material selection—opens up transformative opportunities in on-chip optical communication, sensing, and quantum photonics. Future work will focus on the experimental realization of these concepts and the integration of tunable, EP-based components into scalable photonic circuits, ultimately pushing the frontiers of integrated nanophotonics.
\section*{Supplementary Material}

Supplementary Material is provided to support and extend the results presented in the main text. It contains the following items:

\begin{itemize}
  \item \textbf{Sec.~I — Mode dispersion and modal fields:} full width-dependent plots of $\Re\{n_{\mathrm{eff}}\}$ and $\Im\{n_{\mathrm{eff}}\}$ for the principal guided states (TM$_{\mathrm{LR0}}$, TM$_{\mathrm{SR2}}$, TE$_1$, TE$_2$) and high-resolution field maps (including $|\mathbf{E}|$ and $|E_z|$) that demonstrate spatial coalescence at the exceptional point (see Fig.~S1).
  \item \textbf{Sec.~II — Coupled-mode theory and sensitivity analysis:} derivation of the 2$\times$2 non-Hermitian CMT model, details of the fitting procedure used to extract the parameters, analytic Puiseux (square-root) expansion of the eigenvalue splitting, and quantitative verification of the scaling laws near the EP (see Fig.~S2 and associated fits), suggested measurement approaches (optical transmission, electrical readout from the metal layer, and coherent two-source scattering-matrix reconstruction).
 \end{itemize}

\nocite{*}
\bibliography{sample}

\end{document}